In situ XRD studies of the process dynamics during annealing in cold rolled copper

S.Dey, N.Gayathri*, M.Bhattacharya and P.Mukherjee

Variable Energy Cyclotron Centre, 1/AF Bidhannagar, Kolkata 700064, India

## Abstract

The dynamics of the release of stored energy during annealing along two different crystallographic planes i.e.{111} and {220} in deformed copper have been investigated using in-situ X-ray diffraction measurements at 458K and 473K respectively. The study has been carried out on 50% and 80% rolled Cu sheets. The microstructures of the rolled samples have been characterised using optical microscope and Electron Back Scattered Diffraction measurements. The microstructural parameters were evaluated using Scherrer equation and the modified Rietveld technique from X-ray diffractogram. The stored energy along different planes were determined using the modified Stibitz formula from the X-ray peak broadening and the bulk stored energy has been evaluated using differential scanning calorimetry. The process dynamics of recovery and recrystallization as observed through the release of stored energy has been modelled as a second order and first order process respectively.

*Corresponding author:

Dr. Gayathri N. Banerjee,

Materials Science Studies Division,

Variable Energy Cyclotron Centre

1/AF, Bidhannagar,

Kolkata, 700064

Email: gayathri@vecc.gov.in

Phone: +91-33-23184460

## 1. Introduction:

Metallurgical research in the field of thermomechanical processing is mainly driven by the need of industry. In order to control, improve and optimise the microstructure, grain shape and texture of the final finished product, a detailed understanding of the physical insight underlying the deformation and annealing phenomena is essential [1]. As a result, a major need is to collect the quantitative information involving detailed microstructural evolution during the deformation and annealing processes. The shape of the grains changes with a creation of new grain boundary area by incorporation of the large number of dislocations resulted during the deformation process [1]. The sum of the energy of these dislocations and the new interfaces creating the domains within the grains represents the stored energy of deformation [1]. The increase in the dislocation density is due to the continued trapping of newly created mobile dislocations and their accommodation in the form of the various microstructural features such as elongated grains, cells, domains, subgrains etc. that are the characteristics of the deformed state [2]. The amount of stored energy in the deformed material differs depending upon the configuration of dislocations and their density. In a polycrystalline material, the stored energy of deformation is also highly orientation dependent, since the deformation in the material depends on the slip activity which in turn depends on the grain orientation [3-8]. The stored energy will thus vary from grain to grain depending on its local crystallographic orientation [9,10]. Hence during recrystallization, different crystallographic orientations will show different dynamics of the release of stored energy compared to others because of a more favourable nucleation and/or growth [5,11]. Hence estimation of stored energy as a function of time along different crystallographic planes during annealing of a deformed material will have a significant impact on selecting the process parameters particularly annealing temperature and time, in order to achieve the desired properties at the final finished stage.

The annealing process involves the decrease in stored energy and a corresponding change in the microstructure. Hence the annealing behaviour of a material is not only dependent on the overall stored energy but also on its spatial distribution [1]. Inhomogeneity of stored energy on a local scale will affect the nucleation and the larger scale heterogeneity will influence the growth of the recrystallized grains [1]. Hence, it is of high importance and significance to study the annealing behaviour of a material with different deformations in the light of accumulation of stored energy and distribution of defects. In this paper, we have estimated the evolution of domains and the amount stored energy released at different crystallographic planes during annealing of 50% and 80% cold rolled Cu. Several studies have been done on the annealing of deformed Cu [12-15]. Ph. Gerber et al. [5] carried out studies on evolution of global texture during recrystallization of cold rolled copper after various rolling reductions. G. Mohamed et al. [16] investigated intergranular work hardening state in moderately cold rolled copper and found that the stored energy varied substantially with the initial orientation, leading to a strong variation of the recrystallization temperature. The influence of cold deformation on the stored energy in other alloys have also been studied with the help of neutron diffraction and Electron Back Scattered Diffraction (EBSD) [8,17]. D. Breuer et al. [18] estimated the density of dislocation and its arrangement in plastically deformed copper using X-Ray. T. Rzychon et al. [19] determined the changes in microstructure during compression and oscillatory torsion of deformed copper by XRD line broadening. K. Piekos et al. [6,20] observed stored energy distribution with respect to crystal orientation in polycrystalline copper experimentally as well as using stochastic vertex model of recrystallization. They observed that the release of the stored energy suffered a complicated process and was highly orientation dependent. L. Liu et al. [21] has measured stored energy and recrystallization temperature in polycrystalline copper as a function of rolling strain. C. Deng et al. [7] also estimated the bulk stored energy of cold rolled tantalum by differential scanning calorimetry (DSC) along with the evolution of microstructure. R.D. Doherty et al. [22] described the fundamentals of recrystallization with the understanding of as-deformed state on different metals. D. Mandal et al. [15] studied the effect of grain boundaries on the stored energy at different grain size of the cold worked copper. J. Schamp et al. [23] concluded that the recrystallization is non-homogeneous across the cross-section of the electrolytic tough pitch (ETP) copper wire by measuring the stored energy using DSC. Other studies [2,24] have also been carried out to assess the static and dynamical stored energy on deformed copper. However, the temporal release of stored energy at various deformations and temperature with respect to different crystallographic planes and also the process dynamics involved in releasing the stored energy have not been studied so far.

In this paper, an attempt has been made to characterise the microstructure of electrolytic copper under optical microscope and EBSD in deformed and annealed condition. The evolution of domains during annealing has been studied as a function of time using X-ray Diffraction Line Profile Analysis (XRDLPA). The stored energy has also been estimated qualitatively using modified Stibitz formula [9,25] which is only a first order approximation and the error associated in estimation may be relatively high [26]. High Temperature X-ray Diffraction technique has been used to study the annealing behaviour and the stored energy variation with time in different crystallographic planes. The main aim of this work is to understand the dynamics of the release of stored energy along different crystallographic planes where the absolute value of the stored energy may not be necessarily important. The overall stored energy of the samples has also been estimated by DSC at different deformations. It is to be mentioned that the other important consequences of deformation and/or annealing such as orientation of individual grains or texture have not been addressed here.

## 2. Experimental details

The samples used in this study were electrolytic copper sheets (Table 1) which were deformed at 50% and 80% by cold rolling at room temperature using a roller mill of diameter 16.5 cm at a speed of 20 rpm. Small samples were cut from the rolled sheet by jewellery hacksaw and were mechanically polished using different grade silicon carbide (SiC) papers and finally polished to mirror finish using fine grain diamond paste. Wide Angle X-ray diffraction (XRD) profiles from as received and deformed samples have been recorded by Bruker D8 Advance X-ray diffractometer using $CuK_\alpha$ radiation. A range of 2θ starting from 40° to 100° and a scan step of 0.02° were used. The collection time given for each step was 0.5sec. Instrumental broadening correction was done using a standard defect free Si powder.

For the recrystallization study at the temperatures of 458K and 473K, small pieces of the samples were cut from the rolled sheet and were thinned down to less than 0.2 mm to satisfy the focussing condition of the Brag-Brentano geometry while using the high temperature XRD stage. These samples were polished carefully and mounted on the high temperature XRD stage consisting of a Platinum (Pt) strip heater using a small amount of conducting silver paste. The whole stage was isolated from the surroundings by a vacuum chamber which was maintained at a high vacuum better than $5\times10^{-5}$ mbar during the experiment. For carrying out the time dependent study of annealing at a particular temperature, the sample was heated at a rate of 2 K/sec up to the desired temperature and then the X-ray scan of a characteristic peak e.g. {111}, {220} was taken with different soaking time. To understand the dynamics of the microstructural evolution, it is required to collect the information of the evolution of microstructural features in-situ at a close interval of time. In all cases, a freshly-prepared sample was taken to collect the data for a particular {hkl} plane and also for different temperature. The time for collection of data of each {hkl} plane was chosen after optimizing the peak to background ratio for obtaining an acceptable peak. The 2θ range for the scans was also determined such that the desired amount of back ground points are available on either side of the peak to carry out further analysis. Each peak evolution was followed up to a time till no change could be observed in the integrated intensity values. This indicated that the microstructural evolution (as observed by XRD) has reached its saturation at that particular temperature.

The microstructure of the cold rolled and annealed samples were observed along the plane containing the rolling and transverse direction using optical microscope (Carl Zeiss, Inverted Optical Microscope) and FESEM (Supra 55, Carl Zeiss) equipped with a fully automated EBSD system (Oxford instruments- HKL Channel 5). The Electron Back Scattered Patterns (EBSP) were captured with a beam voltage of 20 kV, at a working distance of 16 mm and a 70° tilt angle. The polished surface was electrochemically etched to obtain a good microstructure. In each case, the sample was mounted with reference to the rolling direction. The microstructure was also observed using EBSD for the samples after the high temperature annealing studies.

The overall stored energy was measured on both the 50% and 80% rolled samples using DSC/TG attachment of the Simultaneous Thermal Analyser (Netzsch STA 449F1). Small pieces (less than 5 mm) of the rolled sheets were used for this study. The experiment was carried out at a heating rate of 40K/min on both the samples up to 873K. After the 1st run, the sample was cooled and reheated at the same rate to acquire the baseline for calculating the quantitative stored energy. Argon gas was used to protect the sample against oxidation during the entire experiment.

## 3. Results and Discussion

### 3.1 Deformation Study

Deformation causes the microstructural changes in various ways. The shape of the grains changes along the direction of deformation and this leads to a large increase in the total grain boundary area [1]. For cubic metals, the main mechanism of deformation is slipping and twinning. The slipping mechanism is mostly predominant for the metals with high stacking fault energy [27]. Though copper has stacking fault energy of $78 mJm^{-2}$ [27], it is still considered to be moderately high [28] and the mechanism of deformation of Cu is primarily by slip followed by twinning. The essential difference between the deformed and the annealed states lies in the dislocation content and their arrangement. It is well understood that the evolution of the microstructure during recovery and recrystallization is determined by the deformation microstructure based on the density, distribution and arrangement of the dislocations [1]. Hence, it is of utmost importance to study the microstructure (grain size and shapes), domains (dislocation cells within the cell blocks), stored energy along different crystallographic planes and also the overall stored energy in the deformed material. In this study we have characterised the deformed samples using various techniques to understand the dislocation and its arrangement and its manifestation in the different length scales.

3.1.1 Deformation Microstructure

We have characterised the grain morphology of the as received and deformed samples by optical microscopy and also by EBSD technique. Fig 1 shows the optical micrographs of all the three samples (as received, 50% rolled and 80% rolled). Both the rolled samples clearly reveal the effect of deformation on the grain morphology. The as-received sample shows clear well defined grains with an average grain size of about 30 μm, which becomes elongated with deformation. The deformation microstructure is more clearly evident in the EBSD micrographs which are shown in Fig (2a, 2b and 2c) for all the three samples. The 50% rolled sample shows a smaller grain size with irregular grain boundaries whereas the 80% rolled sample clearly shows the elongated structure with highly irregular boundaries. The small angle grain boundaries have been generated using the EBSP with a minimum and maximum misorientation angle of $2^0$ and $15^0$ respectively. Using the software, the relative frequency distribution of the misorientation angle between two consecutive EBSP has been evaluated. The EBSD micrograph of the 80% rolled sample (Fig. 2c & 2c') reveals a significantly larger fraction of low misorientation angles as compared to 50% rolled sample (Fig. 2b and 2b'). Another parameter that can be used to understand the effect of deformation is the band contrast (BC) of the EBSP [29]. It is a quality factor derived from the Hough transform that describes the average intensity of the Kikuchi bands with respect to the overall intensity within the EBSP [30]. The values are scaled to a byte range from 0 to 255 (low to high contrast). In general the grain boundaries and deformed regions tend to be darker than the interior of the well recrystallized grains [30]. Similar parameters (such as pattern quality, image quality etc.) defining the quality of the Kikuchi pattern generated in an EBSD have been used recently to understand qualitatively the difference in the microstructure, in the deformed, recovered and recrystallized stages of a sample [17,29-31]. The histogram of the BC values obtained in the three samples are shown in Fig 2a'', 2b'' and 2c'' respectively and it can be clearly seen that the histogram becomes broader and the peak gradually shifts to a lower grey value with increasing deformation.

### 3.1.2 X-ray diffraction studies

The X-ray diffraction pattern of the as received and the deformed samples are shown in Fig 3. It is clearly observed that there is a systematic increase in the intensity of the {220} peak and a decrease in the {111} peak of the 50% and 80% rolled samples respectively compared to the as received sample. A significant broadening of the diffraction peaks of the deformed samples is also observed indicating that the deformation has introduced changes in the internal structure within the grains [1] causing the formation of substructure (domains). During cold rolling, a large number of dislocations are introduced in the material but these are not randomly distributed. These dislocations are accommodated into the material in a variety of ways including creation of internal boundaries. It is seen that in case of copper [1,28] rolled to low strains, the microstructure consists mainly of cells or domains (tangle of dislocations) and also thin plate like microbands.

### 3.1.2(a) Domain size and microstrain evaluation

The X-ray diffraction line profile analysis (XRDLPA) of these samples was carried out to evaluate the microstructural parameters such as size of the domains and microstrain generated due to the deformation. The surface weighted domain size ($D_s$) and the average microstrain values $< \varepsilon_L^2 >^{\frac{1}{2}}$ were obtained by the modified Rietveld technique using LS1 program [32]. The method of analysis has been described in the literature [33]. Fig 4 shows a typical plot of modified Rietveld analysis of the highest deformed sample (80% rolled). The microstructural parameters obtained by this technique for the 50% and 80% rolled samples are shown in Table 2. The volume weighted domain size ($D_v$) along different {hkl} planes was also calculated using Scherrer equation [34] after the appropriate instrumental broadening correction. The integral breadth (β) was calculated considering that the peak profile follows the Lorentzian function. The peaks could not be fitted with a Gaussian or a pseudo-Voigt function which implies that the strain contribution to the broadening is almost negligible. The values of the corrected $D_v$ are also given in Table 2.

It is clearly seen from the Table 2 that both $D_v$and $D_s$ have decreased with increasing deformation resulting in smaller coherent region or domains, but the value of the microstrain is comparatively low and did not change with deformation. Copper being a material with medium stacking fault energy [27,28], the mode of deformation in some grains may be accommodated by two mechanisms, slipping and twinning. It is observed that the relative shift in 2θ is not significant in case of both 50% and 80% rolled samples as compared to as received sample. Hence, it may be considered that the stacking fault is almost absent even after 80% deformation. With the increase in strain, the dislocation initially forms the cell structure and subsequently the deformation is accommodated by twinning [1,30], the amount of which increases with increasing deformation. However, we could not observe any deformation twins in FESEM as these may be extremely fine. Humphrey et.al.[1] have reported that the thickness of these twins are in the range of 0.2-0.3nm.

3.1.2(b) Calculation of Stored Energy along crystallographic planes

Rajmohan et.al. [9] have used a modified Stibitz [25] formula for calculating the stored energy along different crystallographic orientations [9] using direction dependent elastic modulus [$E_{hkl}$] & Poisson's ratio [$\nu_{hkl}$] as given below.

$$E_{<hkl>} = \frac{3}{2} Y_{<hkl>} \frac{\left(\frac{\Delta d}{d}\right)^2}{(1+2\nu_{<hkl>}{}^2)} \tag{1}$$

where, $Y_{<hkl>}$ is the Youngs modulus and $\nu_{<hkl>}$is the Poisson's ratio which are both direction dependent. The relative change in the lattice spacing $\left(\frac{\Delta d}{d}\right)$ can be obtained from the broadening of the diffraction peaks from the following relation:

$$\left(\frac{\Delta d}{d}\right) = \frac{\beta}{2\tan\theta} \tag{2}$$

where, $\beta$ is the integral breadth after instrumental broadening correction at the Bragg angle $2\theta$ as explained in 3.1.2(a).

The values of the stored energy along the planes {111}, {200}, {220} and {311} have been estimated using the direction dependent $Y_{<hkl>}$ and $\nu_{<hkl>}$ and are presented in Table-3 as a function of deformation. It is evident from Table 3 that the stored energy is highly orientation dependent and the value of stored energy is more in the highest density planes i.e. {111} as compared to the other planes even in the as received samples. The stored energy has increased in all the planes with increasing deformation with the change being maximum in the {111} plane. The other planes show a relatively lower increase in the stored energy. {111} being the slip plane can accommodate more dislocations and hence show a greater increase in stored energy as compared to other planes. It should be noted that the error in full width at half maximum value was high for the peaks with low signal to noise ratio ({311} peak) and also low intensity values.

### 3.1.3 Stored energy by Differential Scanning Calorimetry

The stored energy provides the driving force for recovery and recrystallization [1,25]. Hence it is necessary to quantify the overall stored energy in both the samples which will help to determine the annealing temperature. Fig 5shows the DSC curves of the 50% and 80% rolled samples heated at a rate of 40K/min after the baseline correction. It is clearly seen that the peaks are obtained at temperature of 592K and 552K for 50% and 80% rolled samples respectively. The area under the curve of the peak gives a measure of the stored energy. It is found to be 38 J/mole and 46 J /mole for 50% and 80% rolled samples as seen in Fig.5. These values corroborate with the studies done by L. Liu et. al. [21] as a function of strain in pure copper.

It is evident that the stored energy of the 80%rolled sample is larger as compared to the 50% rolled sample. Moreover, the release of the stored energy is found to occur at a lower temperature for the 80% deformation. It is well known that the stored energy causes the recovery and the recrystallization process to occur at a lower temperature [35].This may be attributed to the fact that smaller activation energy is required to release the overall stored energy for the 80% rolled sample, causing the process to start at a lower temperature.

## 4. Study of Recovery and Recrystallization

When a deformed material is annealed, recovery and recrystallization occur resulting in the formation of defect free grains and thus restoring the property of the material to a significant extent [2,27]. Both the recovery and recrystallization are competing processes and both are driven by the stored energy in the deformed sample [1]. During the recovery process, the microstructural changes are very subtle and occur on a small scale while new defect free grains are nucleated and subsequently grow during the recrystallization process [1]. A division between recovery and recrystallization is thus difficult to define; since recovery mechanism plays a crucial role in nucleating recrystallization [22]. As recovery lowers the driving force for recrystallization, a significant amount of prior recovery may in turn influence the nature and kinetics of recrystallization [36-38]. The evolution of recovery and recrystallization processes in a material depends on the extent of deformation (stored energy in the sample) and also on the nature of the material ie, the stacking fault energy, impurities etc [1]. As observed in the last section, the accumulation of stored energy is highly orientation dependent and thus it is expected that the release of the stored energy along different crystallographic orientations will show a different behaviour as a function of time. A comprehensive study has been made in-situ to follow the release of stored energy with time at a particular temperature to understand the dynamics of recovery and recrystallization.

The temperatures for the dynamic study were chosen based on the information from the reported literature [5]. Ph. Gerber et.al. [5] have observed that for 90% and 70% rolled pure Cu sample recrystallization of pure Cu starts at around 398 K and 448 K and ends at 473 K and 573 K respectively. Hence, in this study, the annealing was done at two different temperatures (458 K and 473 K) to understand the dynamics.

The major slip system in copper is {111}<110> [27]. In this study, during the deformation process (rolling) the {220} planes are found to be oriented along the rolling plane which is evident from the increase in the peak intensity as seen in Fig 3.It is known that during the rolling process, if the compressive stress is more than the longitudinal stress, then the majority of {220} planes tend to orient towards the rolling plane [39].This is because, the {111}<110> slip system becomes active tending to align the {111} planes perpendicular to the rolling plane and since <110> is the slip direction, the corresponding {110} planes tend to align parallel to the plane containing the rolling and the transverse directions. As a result, the two major crystallographic planes which show predominant changes in the values of the integrated intensity during the deformation process are the {111} and the {220} planes. Hence the annealing experiment using high temperature XRD has been carried out on these two crystallographic planes only.

### 4.1 In-situ dynamic studies of recovery and recrystallization using High Temperature XRD

#### 4.1.1 Variation of domain size in different planes

The single peak analysis of the XRD peaks of {111} and {220} set of planes have been performed to characterise the in-situ evolution of microstructure with time at different elevated temperatures. The true integral breadth of {111} and {220} peaks as a function of time were evaluated after correcting the Debye Waller factor and instrumental broadening at 458K and 473K. The volume weighted domain size for {111} and {220} was estimated using the Scherrer's equation [34] and plotted as a function of time. Fig. 6(a) and 6(b) reveal that at two different temperatures, a distinct variation in the size of the domains in {111} and {220} planes are observed as a function of time at the initial stage for both the deformed samples. The domain size increased rapidly followed by a saturation for {111} and {220} peaks at 473K. But the rise in the domain size is observed to be gradual following a straight line at the initial stage and then saturates with time at an annealing temperature of 458K. In the heavily deformed sample, recovery occurs prior to recrystallization primarily with the changes of dislocation structure in the material. It is to be noted that recovery and recrystallization are competing processes and both are driven by the stored energy of the deformed state [1]. Initially there is a rearrangement of dislocations followed by annihilation which causes increase in the domain size. This may be considered as the recovery stage and the onset of saturation in the domain size as the start of recrystallization stage. The variation in the profile of the evolution of domain size at two different temperature (Fig 6(a) and 6(b)) may have occurred due to the difference in the amount of thermal energy supplied to the deformed material. Hence, the increase in annealing temperature causes a significant amount of prior recovery at a much lower time.

4.1.2 Variation of stored energy in different planes

Fig 7 shows the variation of the stored energy with time as calculated using equation (1) for the 80% rolled sample from the diffraction peak broadening along the {111} planes in a semi-log plot. The variation along {111} planes is monotonic (similar to an exponential decay) at two different temperatures. It is worth noting that the saturation values of the stored energy at the end of the experiments for both the temperatures are same (within the error bar) which gives a good confidence in the experimental procedure. Fig 8(a) and 8(b) shows the stored energy release with time along the {220} planes for the 50% and the 80%rolled samples at the temperature458K and 473K respectively. It can be clearly seen that the nature of the variation of the SE with time in the {220} planes is quite different compared to that of {111}. The curve shows two distinct regions: a smaller decrease in the stored energy in the early stage (stage I) and then an exponential decay (stage II), almost similar to the behaviour of {111} set of planes (Fig. 7). The time corresponding to the change-over between the two kinetics is significantly different in the two deformed samples (marked by arrows in Fig.8(a) and 8(b)). The change in the process dynamics is found to occur at a lower time for 80% deformation as compared to 50%, which is expected. In the initial stage, recovery occurs by the rearrangement and annihilation of dislocations and the stored energy vs time curve shows a different time dependence compared to that in the second stage which arises due to the initiation of recrystallization processes seen Fig. 8. Since, the amount of stored energy is the driving force of recrystallization, the onset of recrystallization occurs at a lower time for higher deformation (80% deformation). This is also evident from the clear signature of the initial increase in intensity for {220} peaks due to recovery followed by a sharp decrease due to the formation of recrystallized texture as seen in Fig 9. This phenomenon could not be observed for the {111} planes as the release of stored energy and the dynamic process of annealing is so fast that it could not be followed within the laboratory time scale. The faster dynamics in the {111} can be explained by considering the fact that this plane has the highest the planar density of atoms and the initial stored energy is highest (Table 3) indicating that this plane can accommodate more dislocations during the deformation process and subsequently on annealing, the stored energy is released at a much faster rate.

These dynamic processes of annealing responsible for release of the stored energy can be mathematically modelled in the following way. Recovery and recrystallization are the primary process involved in the change in the stored energy of the deformed material. Recovery involves the movement of the dislocations and subsequently their annihilation causing a decrease in the stored energy, and recrystallization involves the process of nucleation and growth of new defect free grains which is also again driven by the stored energy in the material. The kinetics of these two processes will hence determine the overall variation of the stored energy as a function of time. In a medium stacking fault energy material, the mobility of the dislocation is reduced [40] and hence the movement of the dislocations can thus be considered to be restricted to quasi two dimension and can be modelled as a second order process. The recrystallization on the other hand is restricted to the boundaries of the grains and the interfaces, and hence can be modelled as a first order process [1,41]. As we clearly see from the experimental results above, the kinetics of the release of stored energy is different along the {111} and {220} crystallographic planes in the experimental duration. Hence, it can be conjectured that the time scale of the two processes involved in the release of the stored energy is different along the two planes. This is expected because, {111} set of planes have a higher stored energy and would thus have a lower activation energy compared to the {220} set of planes. The stored energy variation is thus fitted to the following equation.

$$SE(t) = SE_{t=\infty} + Ae^{-\left(\frac{t+t0}{\tau_1}\right)^2} + Be^{-\left(\frac{t+t0}{\tau_2}\right)} \qquad (3)$$

where, $\tau 1$ and $\tau 2$ are the time constants of the second order (recovery) and the first order (recrystallization) process respectively. A time '$t0$' has been included in the two terms to account for the experimental delay time (required to be introduced to account for the time during which both the processes have already began before the start of the experimental observation). Coefficients A and B give the relative contribution of recovery and recrystallization process to the dynamics. The stored energy variation has been fitted using this equation and the values obtained are tabulated in Table 4.The fitting curves are superposed on the experimental data in Fig 7, 8(a) and 8(b).It can be seen from the Table 4 that the time constants of the two processes are varying systematically with temperature as well as the deformation. In the 80% rolled sample, the time constants of the recovery and recrystallization processes are much less along the {111} plane compared to the {220} plane. In fact, at T=473K, the curve could be fitted with only the first order process with a small time constant of 423 secs. Along the {220} plane, the 80% rolled sample has lower time constants compared to the 50% rolled sample. This is also expected since the 80% rolled sample has a higher stored energy thus enabling the recovery and also the recrystallization process to start at a earlier time compared to that of the 50% rolled sample. From table 4 it can be seen that for the 50% rolled sample, annealed at 458K, the time constants of the two processes are quite large and is comparable to the maximum experimental time. It can also be seen from Fig.8(b) that the release of the stored energy is not complete in the sample subjected to 458K within the experimental time. This can also be corroborated with the EBSP analysis and the misorientation angle distribution of the same sample as discussed below. The large differences in the time constants along the {111} and {220} planes clearly indicate that there is a marked difference in the kinetics of the two processes ie., recovery and recrystallization along the different crystallographic planes and the activation energy of the processes is also different in the different planes.

#### 4.1.3 Microstructure of the annealed samples

The microstructure of the annealed samples using EBSP shows clear indication of well recrystallized grains as shown in Fig.10(a) and Fig.10(b). It is evident from the observations that the irregular shaped boundaries as seen in Fig.2(b) and 2(c) have changed to sharper boundaries with larger misorientation among the grains. This is especially more significant in the 80% rolled sample. This also corroborates with the XRD peaks [Fig.11] which show clear changes in the peak intensity of {200} and {220} planes indicating the formation of recrystallized texture. It can be seen that for the 80% rolled sample, both the misorientation angle distribution (Fig 10(b′), which shows a significant number of high misorientation angle boundaries) and the band contrast distribution (Fig 10(b′′) which shows a peak shift towards the high grey values and a much sharper distribution function) clearly shows that new defect free grains have formed. In the case of the 50% rolled sample, the misorientation angle distribution (Fig 10(a′)) and the band contrast (Fig 10(a′′)) clearly indicates that significant recrystallization has not taken place; there are formation of new grains with lower misorientation angle and only a slight shift of the band contrast distribution function to higher grey values.

## 5. Conclusion

Detailed characterisation of deformed materials has been carried out with respect to microstructure (grain size, domain size, microstrain), stored energy (overall and along specific planes), temporal evolution of microstructure and release of stored energy at different temperatures to study the annealing phenomenon. The deformed Cu samples have been used to understand the kinetics of the release of stored energy along different crystallographic planes during the annealing process. The microstructure was characterised using EBSP which revealed irregular grain boundaries with increasing deformation. The microstructural parameters as evaluated using XRDLPA technique showed a decrease in domain size whereas the microstrain remained almost constant. The overall stored energy was found to increase with deformation. The stored energy along the different crystallographic planes was evaluated using the modified Stibitz formula and it was seen that the accumulation of stored energy is much higher in {111} set of planes compared to the other planes. The kinetics of the release of the stored energy along the {111} and {220} set of planes were followed using in-situ XRD measurements carried out at 458K and 473K on the deformed samples. The process dynamics of the recovery and the recrystallization could be explained clearly from the observed time dependent variation of the stored energy in the {220} set of planes. The two processes were found to follow second order and first order dynamics respectively.

Table 1. Composition of the Copper sample

| Element | O | Ag | S | Fe | Ni | Sn | As | Pb | Sb | Cu |
|---|---|---|---|---|---|---|---|---|---|---|
| Amount in ppm | 270 | 25 | 14 | 10 | 8 | 5 | 4 | 3 | 3 | Bal. |

Table2. Volume weighted domain size (calculated using Scherrer formula) of different {hkl} planes and surface weighted domain size and microstrain (calculated using modified Rietveld technique) of the as received and deformed samples

| Sample | Domain size of {111} at RT (nm) | Domain size of {200} at RT (nm) | Domain size of {220} at RT (nm) | Domain size of {311} at RT (nm) | Domain size $D_s$ (nm) | Microstrain |
|---|---|---|---|---|---|---|
| As received | 48.73 | 27.40 | 33.86 | 25.34 | ---- | ---- |
| 50%rolled | 35.53 | 27.96 | 26.96 | 19.63 | 29.4 | $8.6x10^{-4}$ |
| 80%rolled | 32.76 | 21.39 | 26.15 | 16.32 | 24.5 | $8x10^{-4}$ |

Table 3.Stored Energy of different {hkl} planes of the As-received, 50% and 80% rolled samples

| Sample | SE of {111} at RT (J/Mole) | SE of {200} at RT (J/Mole) | SE of {220} at RT (J/Mole) | SE of {311} at RT (J/Mole) |
|---|---|---|---|---|
| As received | 10.48 | 7.48 | 5.22 | 4.79 |
| 50% rolled | 19.92 | 7.15 | 8.09 | 7.73 |
| 80% rolled | 23.17 | 12.24 | 8.63 | 11.35 |

Table 4. Parameters obtained by fitting eq. 3 to the observed stored energy variation as a function of time.

| Sample | Temperature (K) | Peak | $SE_{t=\infty}$ (J/mole) | t0 (secs) | τ1 (secs) | τ2 (secs) | A (J/mole) | B (J/mole) |
|---|---|---|---|---|---|---|---|---|
| 80% rolled | 458 | {111} | 2.3 | 1200 | 704 | 1420 | 409.3 | 23.7 |
| | 473 | {111} | 2.0 | 577 | - | 423 | 0 | 23.5 |
| | 458 | {220} | 0.6 | 1200 | 7565 | 9784 | 3.8 | 4.4 |
| | 473 | {220} | 0.7 | 577 | 2461 | 2657 | 3.5 | 6.6 |
| 50% rolled | 458 | {220} | 0.9 | 1200 | 12160 | 17500 | 3.3 | 4.6 |
| | 473 | {220} | 1.6 | 577 | 1729 | 3615 | 1.3 | 10.6 |

Figure Captions

Fig. 1. Optical Micrographs of (a) As received (b) 50% rolled and (c) 80% rolled copper

Fig. 2. EBSD Micrograph of (a) As-Received, (b) 50% rolled and (c) 80% rolled; Frequency distribution of misoriented angles of (a′) As-Received, (b′) 50% rolled and (c′) 80% rolled ; Frequency distribution of band contrast (a′′) As-Received, (b′′) 50% rolled and (c′′) 80% rolled

Fig. 3. X-ray diffraction patterns of the samples.

Fig. 4. Typical plot of modified rietveld analysis of the 80% rolled sample

Fig. 5.Typical DSC curves of 50% rolled and 80% rolled samples

Fig. 6. Domain Size variation on (a) {111} planes and (b) {220} planes at two temperatures 458 K and 473 K.

Fig. 7. Stored Energy variation in {111} planes at two temperatures 458 K and 473 K on 80% rolled Cu. The red line shows the fit curves according to eqn 3.

Fig. 8. Stored Energy variation for {220} plane at two temperatures 458 K and 473 K on (a) 50% rolled and (b) 80% rolled. The red line shows the fit of the data according to eqn 3.

Fig. 9.Stored Energy and peak Intensity variation for {220} plane at temperature 458 K on 80% rolled Cu.

Fig.10. EBSD Micrograph of annealed samples (a) 50% rolled and (b) 80% rolled; Frequency distribution of misorientation angle of the annealed samples (a′) 50% rolled and (b′) 80% rolled; Frequency distribution of band contrast of the annealed samples (a′′) 50% rolled and (b′′) 80% rolled.

Fig. 11 X-ray diffraction patterns of the 80% rolled sample before and after annealing. (Arrows indicate the decrease in the (220) intensity and the increase in the (200) intensity).

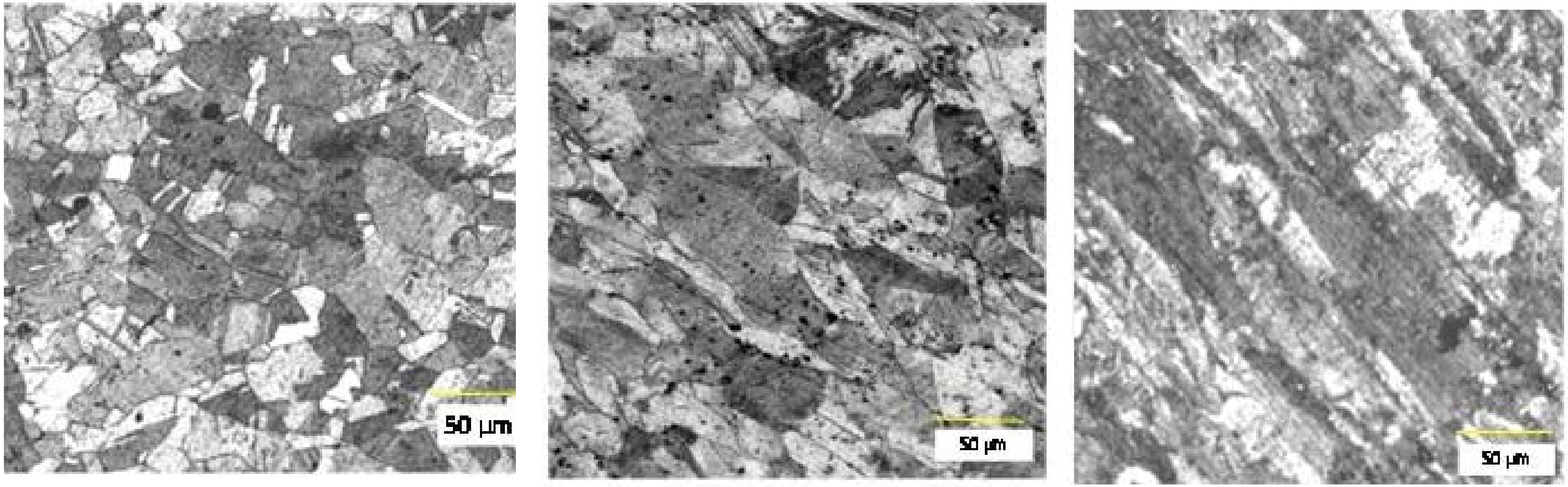

(a) (b) (c)

Fig. 1. Optical Micrographs of (a) As received (b) 50% rolled and (c) 80% rolled copper

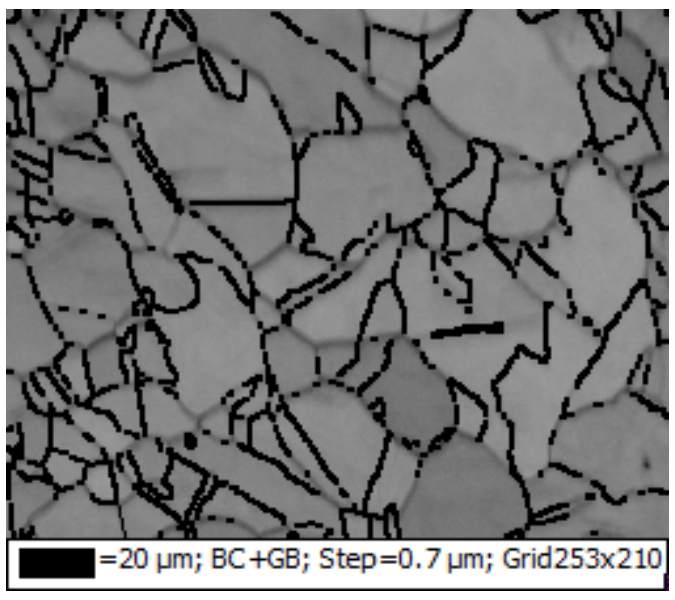

Fig. 2 (a)

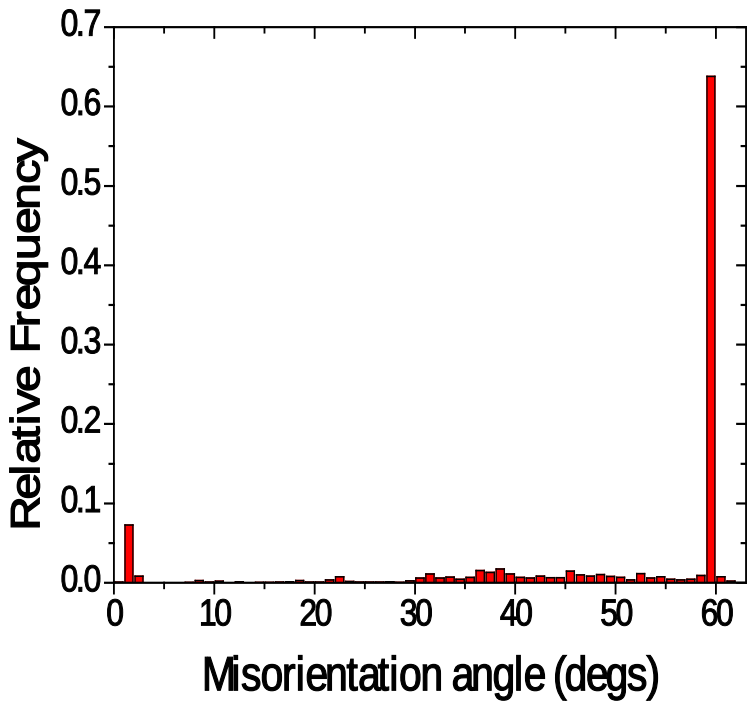

Fig. 2 (a´)

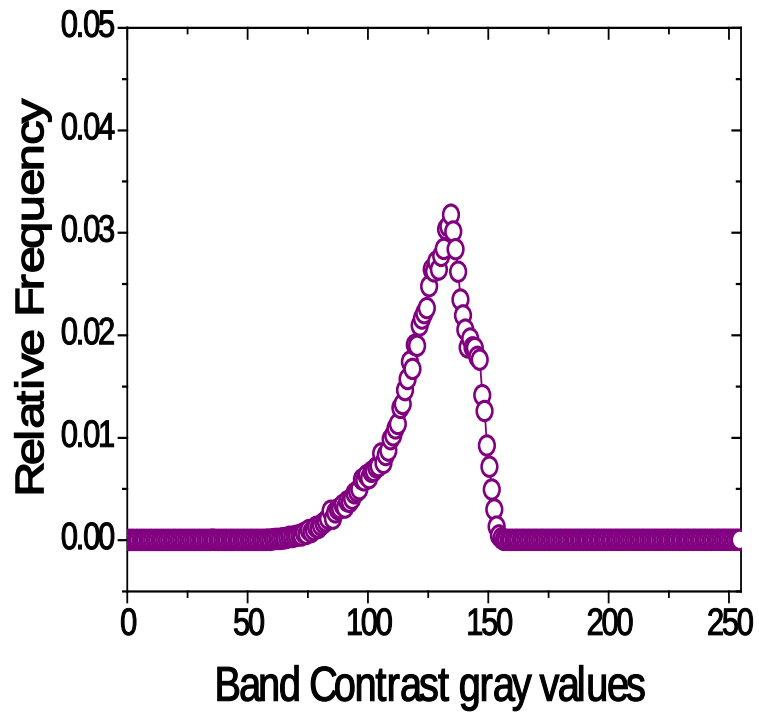

Fig. 2 (a´´)

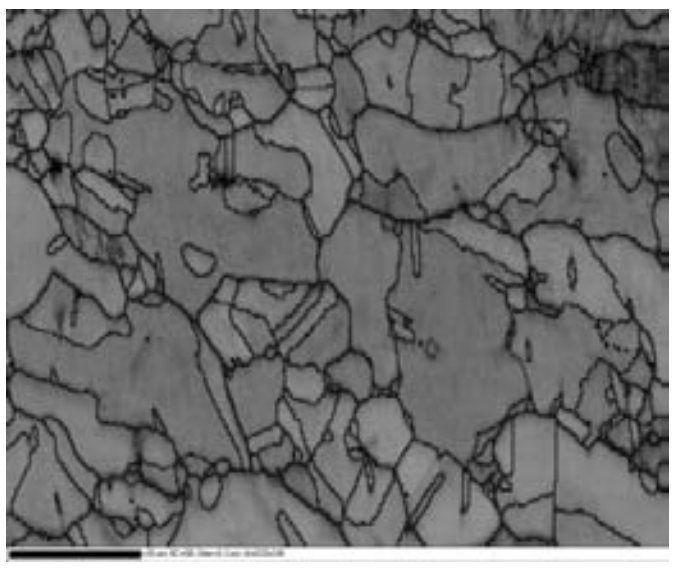

Fig. 2 (b)

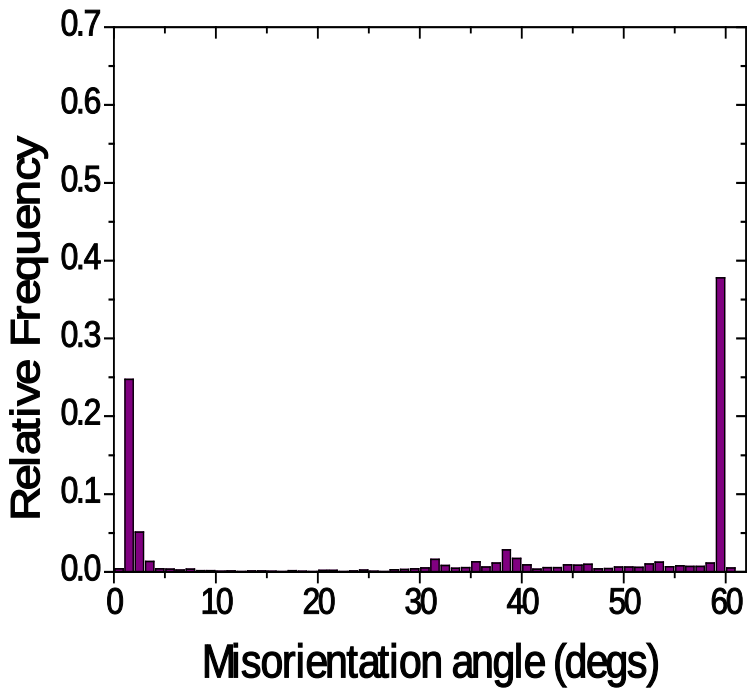

Fig. 2 (b´)

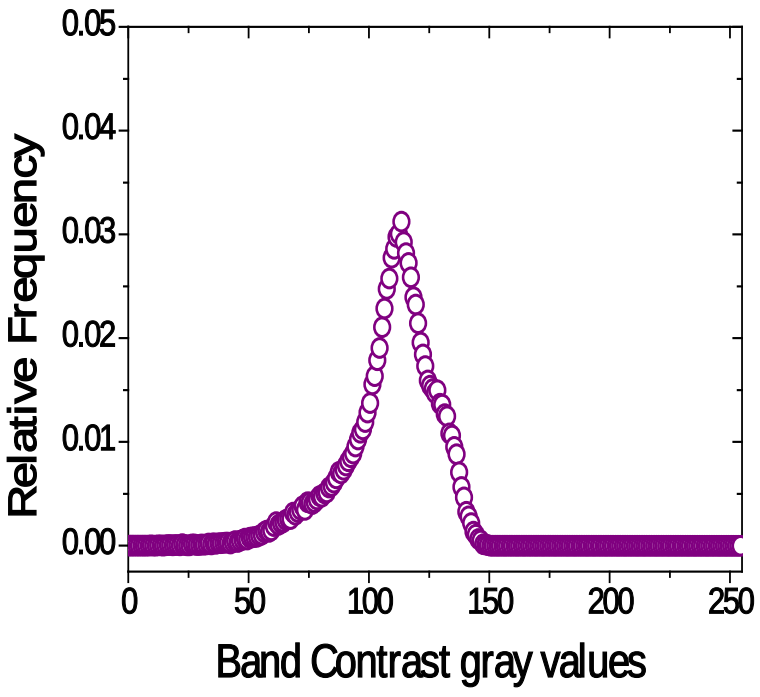

Fig. 2 (b´´)

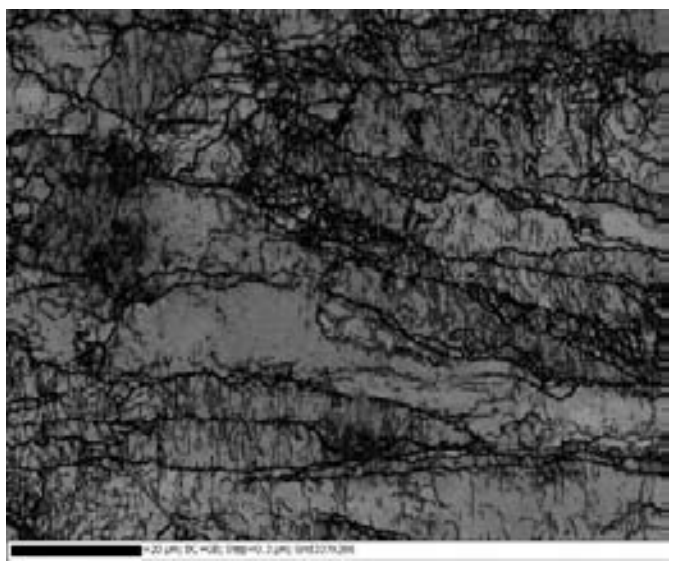

Fig. 2 (c)

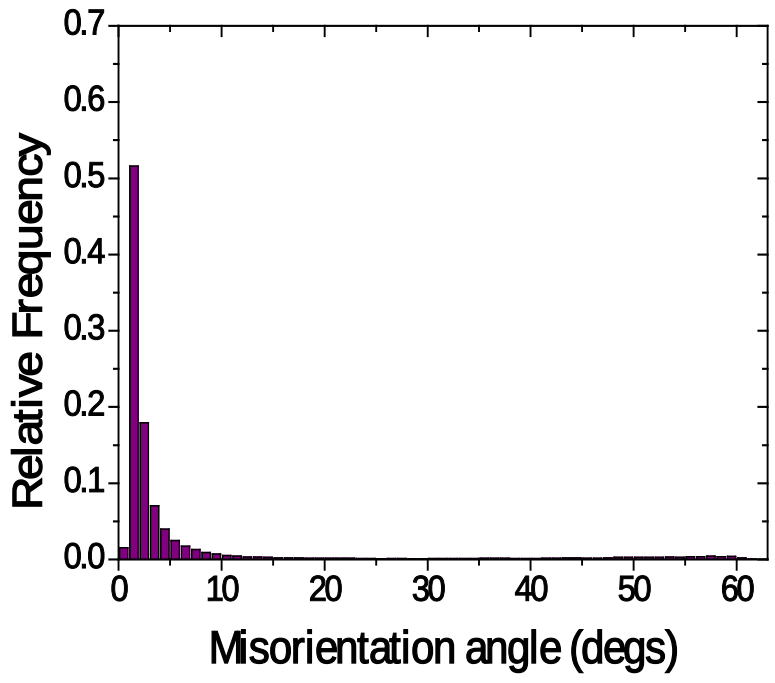

Fig. 2 (c´)

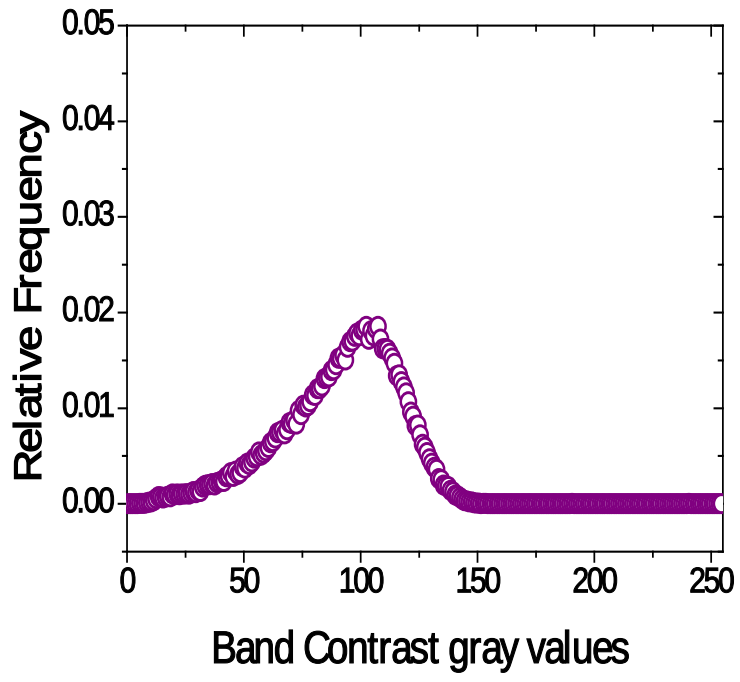

Fig. 2 (c´´)

Fig. 2. EBSD Micrograph of (a) As-Received, (b) 50% rolled and (c) 80% rolled ; Frequency distribution of misoriented angles of (a´) As-Received, (b´) 50% rolled and (c´) 80% rolled ; Frequency distribution of band contrast (a´´) As-Received, (b´´) 50% rolled and (c´´) 80% rolled

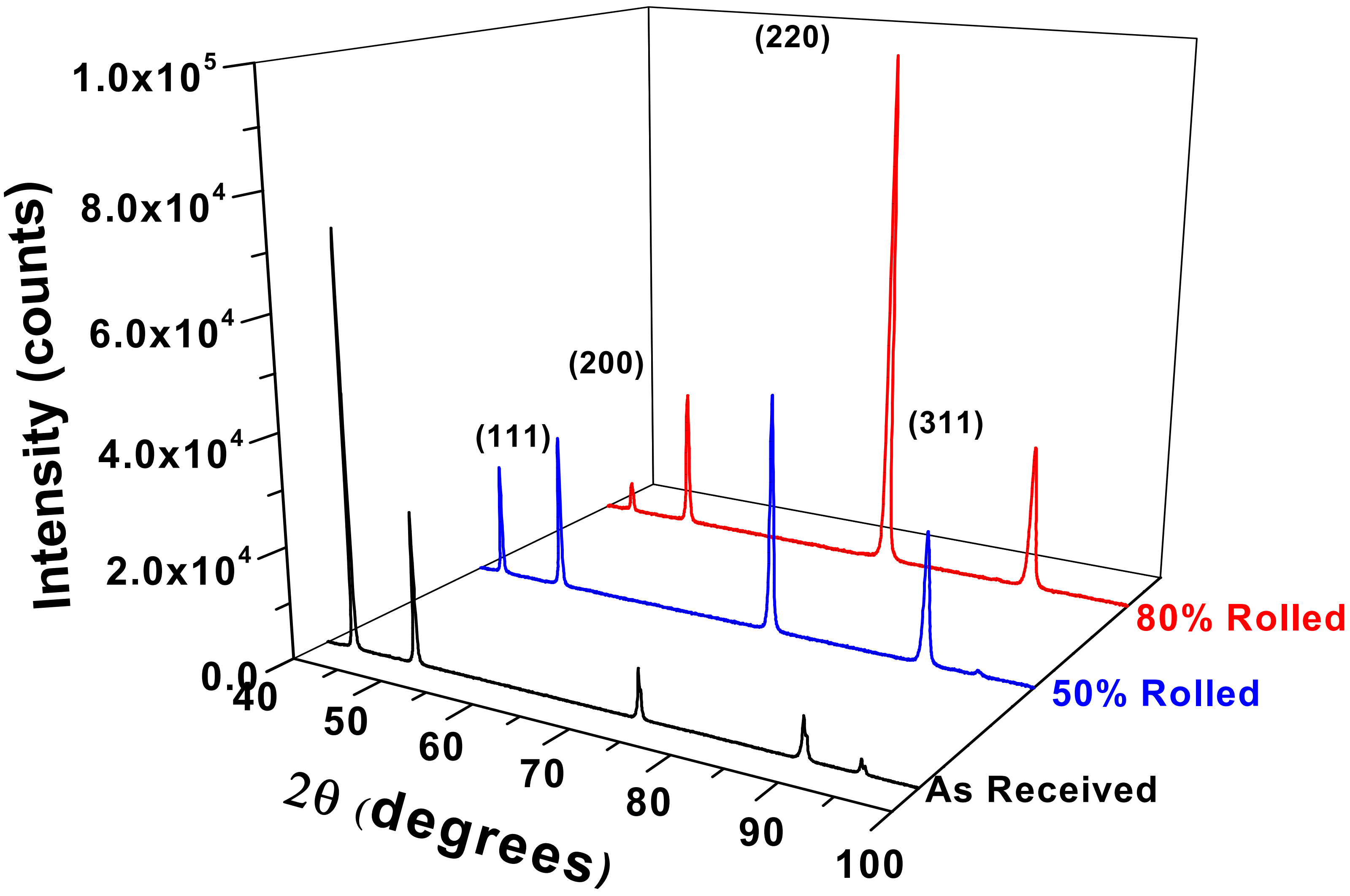

Fig. 3. X-ray diffraction patterns of the samples.

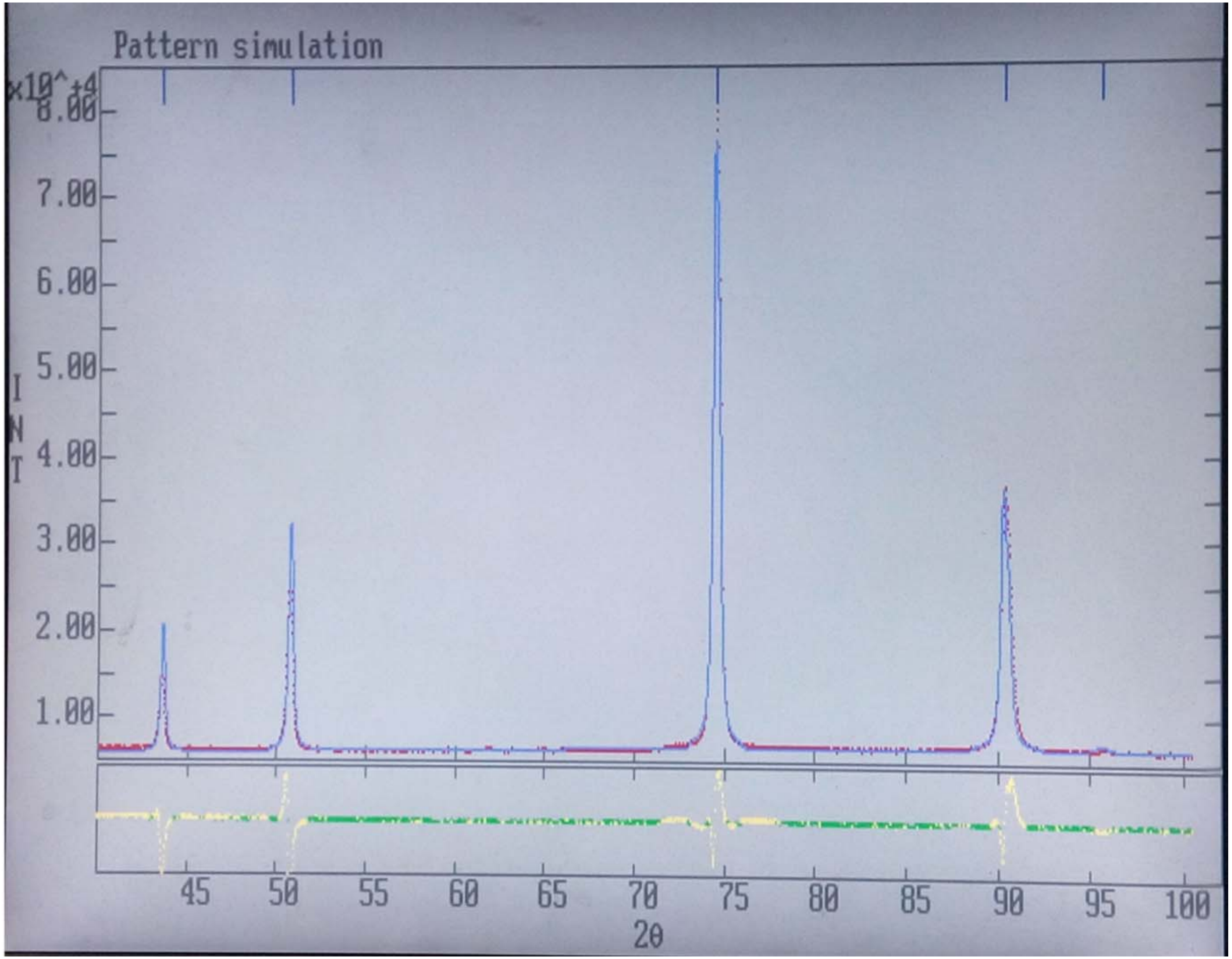

Fig. 4 Typical plot of modified rietveld analysis of the 80% rolled sample

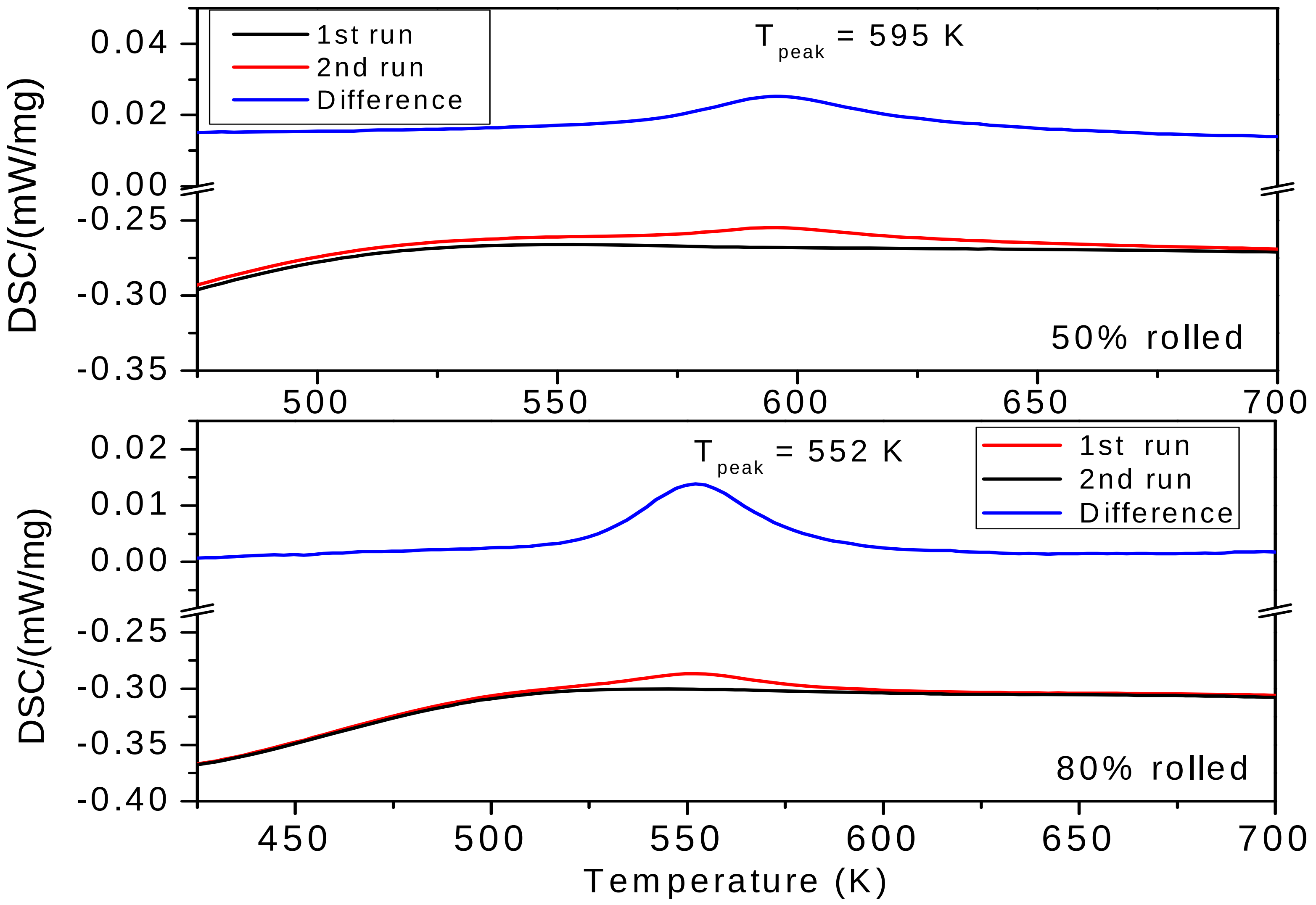

Fig. 5. Typical DSC curves of 50% rolled and 80% rolled samples

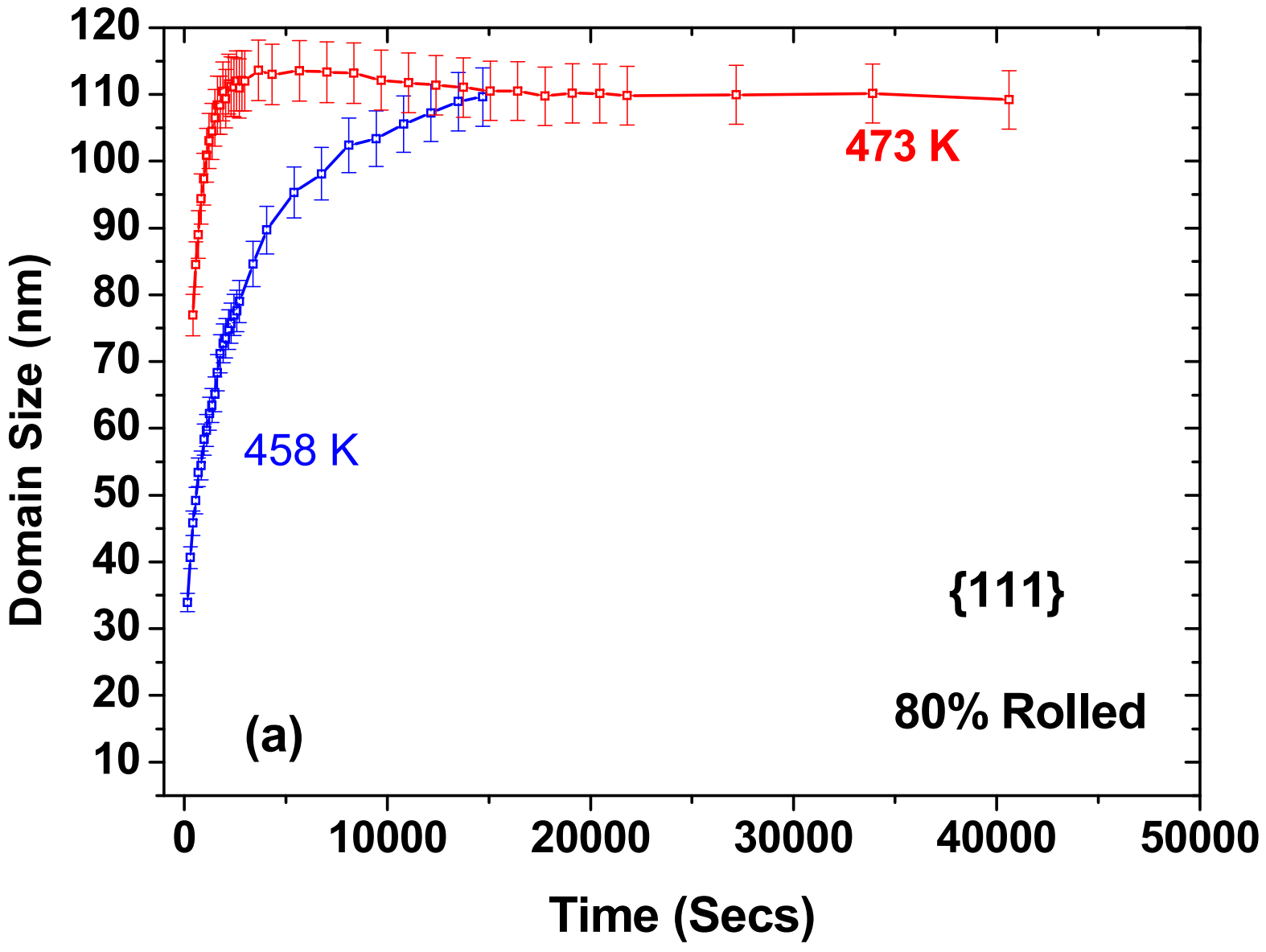

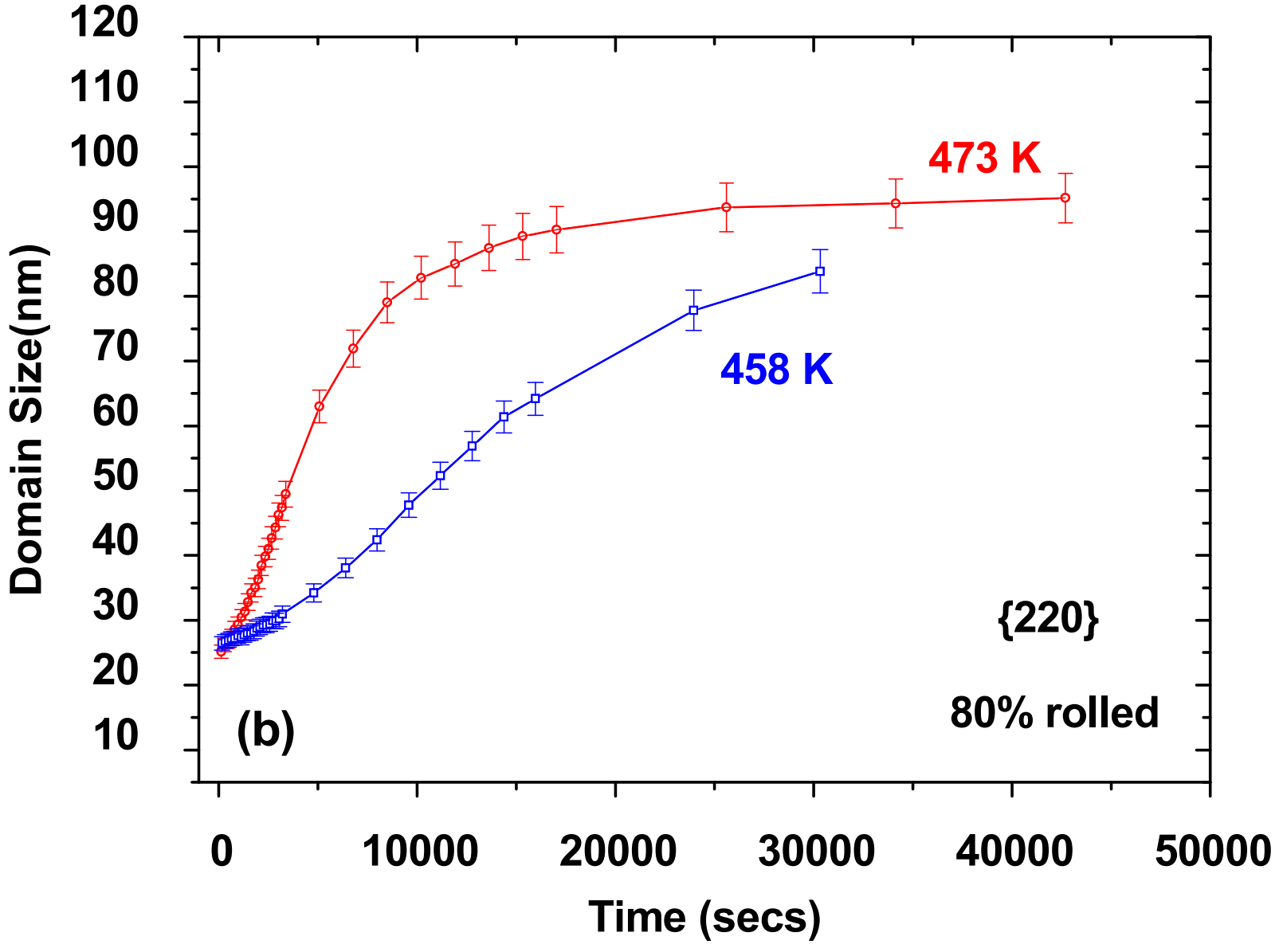

Fig. 6. Domain Size variation on (a) {111} planes and (b) {220} planes at two temperatures 458 K and 473 K

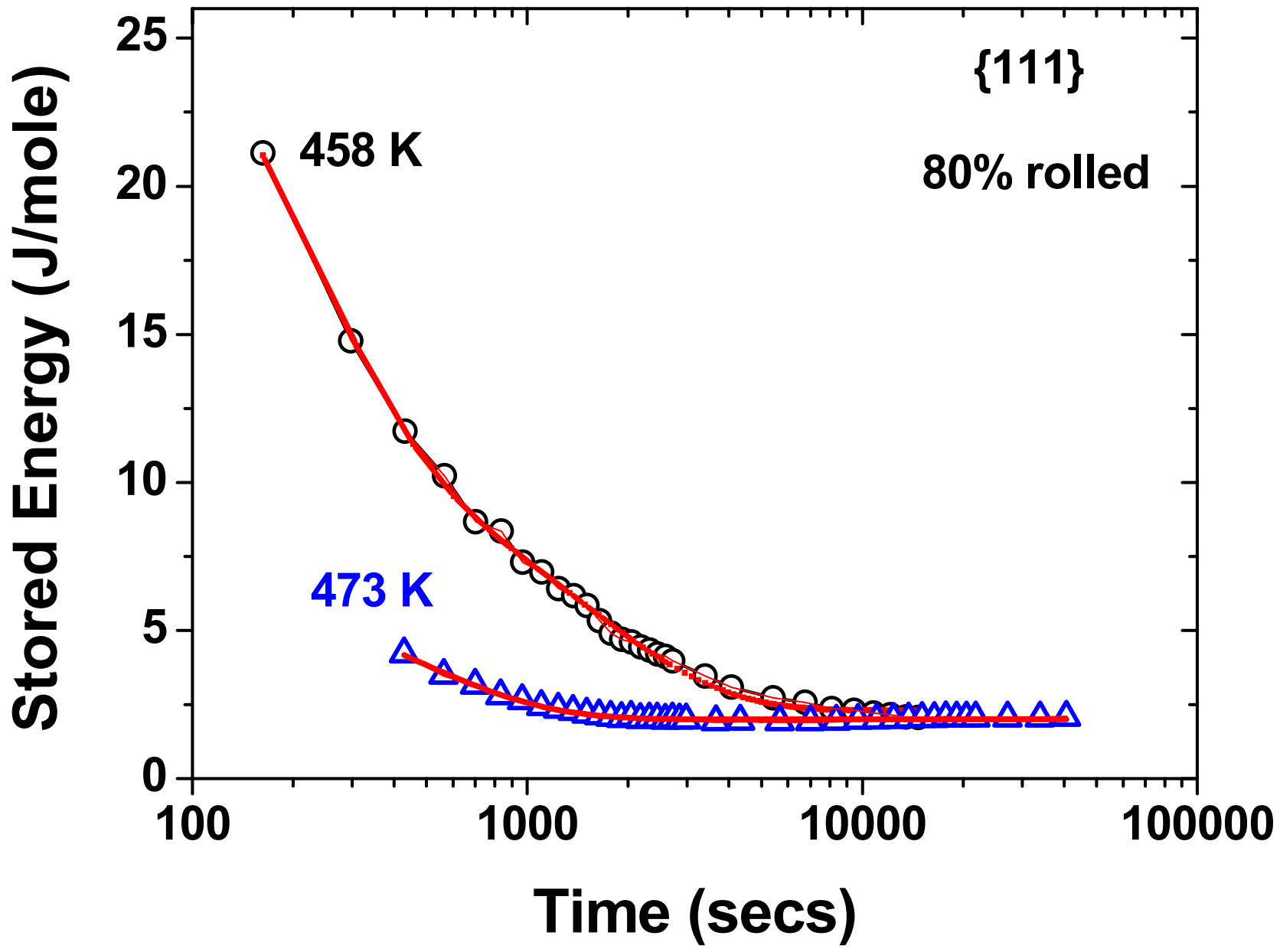

Fig. 7. Stored Energy variation in {111} planes at two temperatures 458 K and 473 K on 80% rolled Cu. The red line shows the fit curves according to eqn 3.

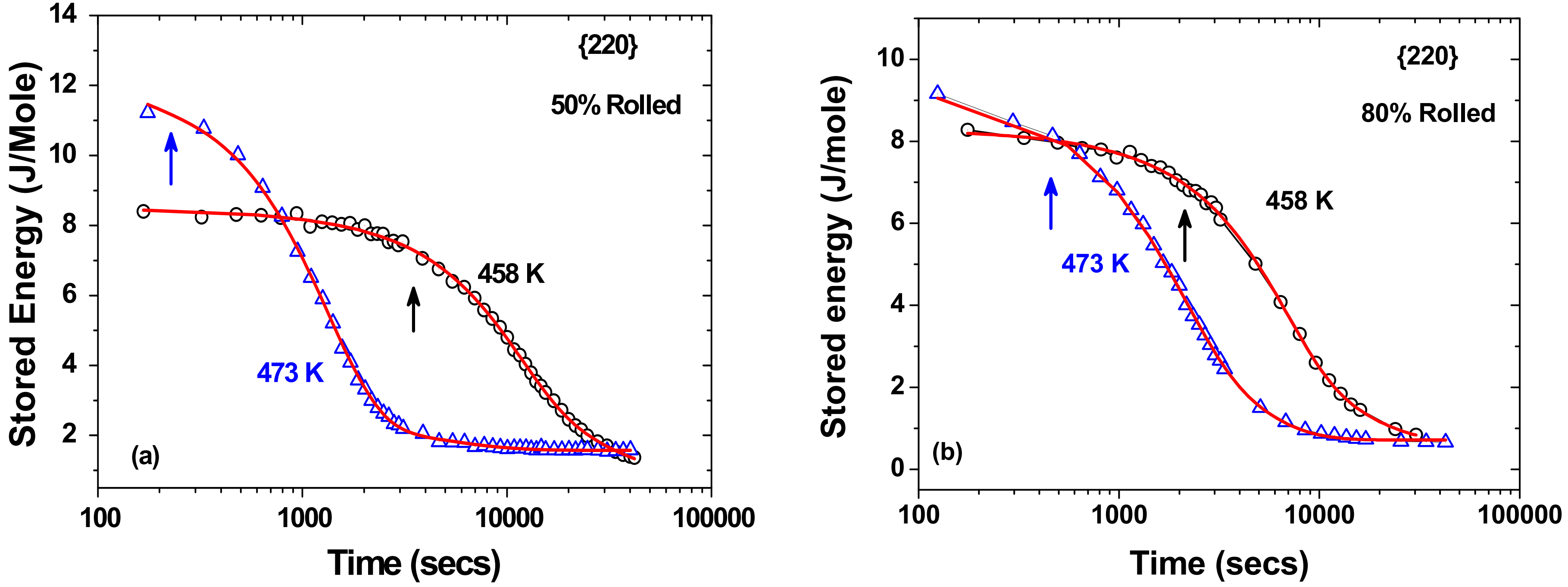

Fig. 8. Stored Energy variation for {220} plane at two temperatures 458 K and 473 K on (a) 50% rolled and (b) 80% rolled. The red line shows the fit of the data according to eqn 3

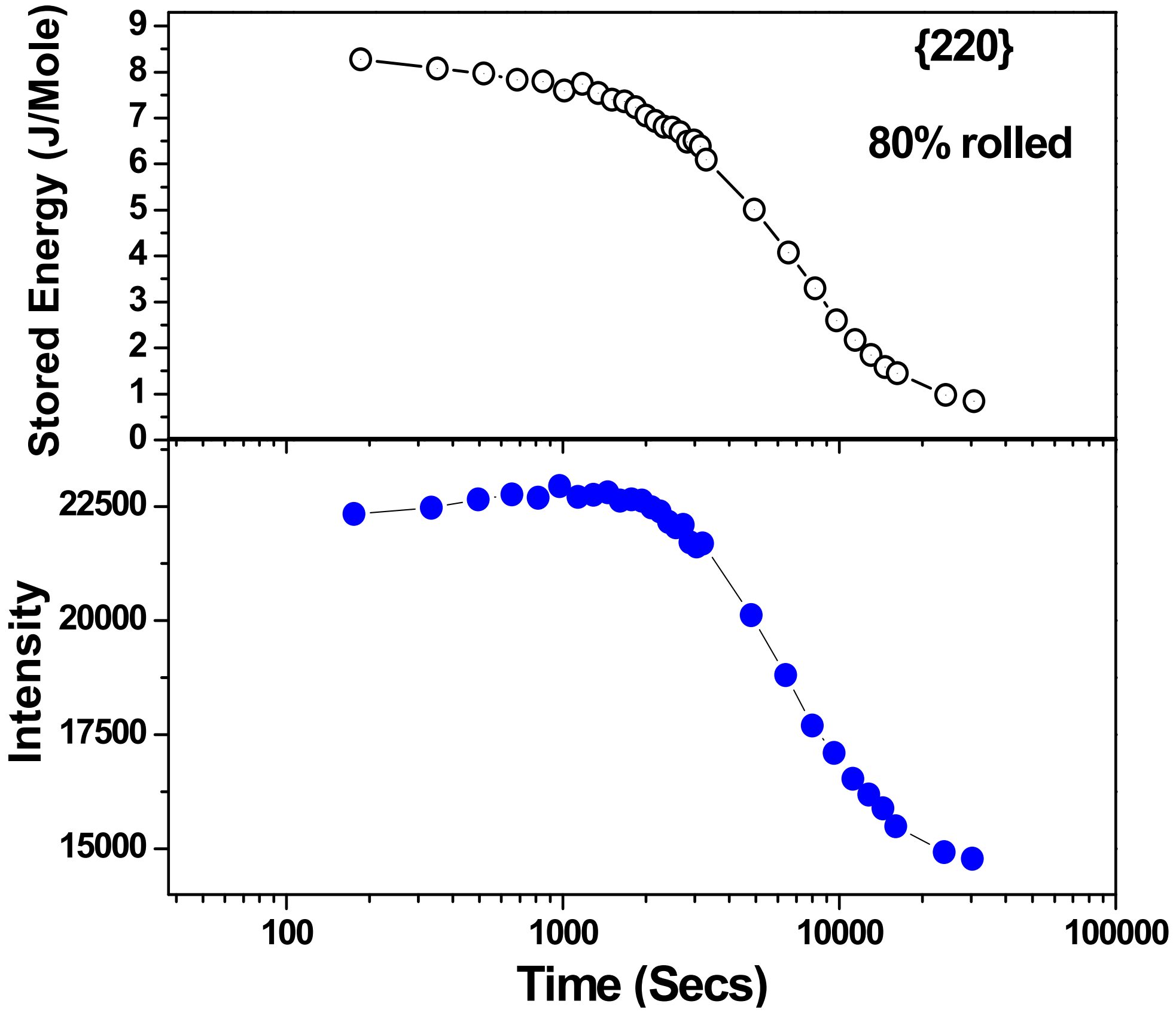

Fig. 9. Stored Energy and Peak Intensity variation for {220} plane at temperature 458 K on 80% rolled Cu

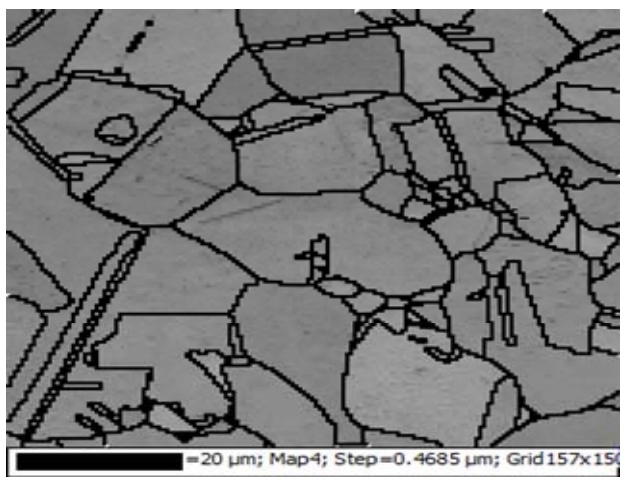

Fig. 10(a)

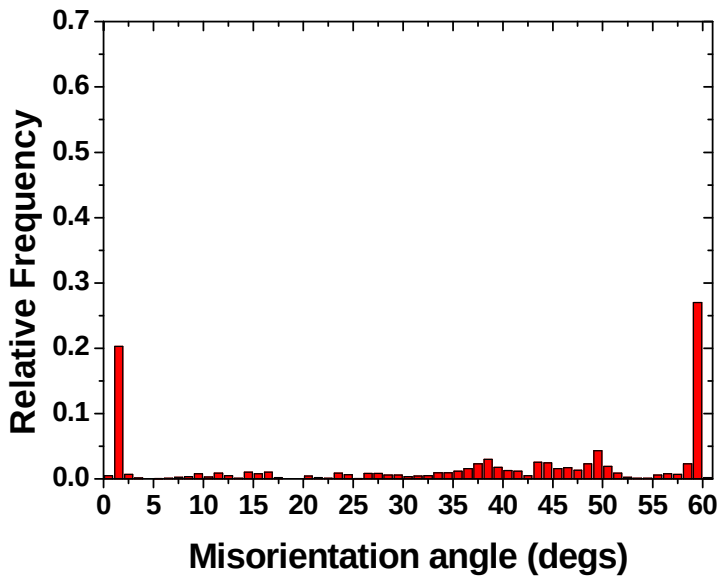

Fig. 10(a´)

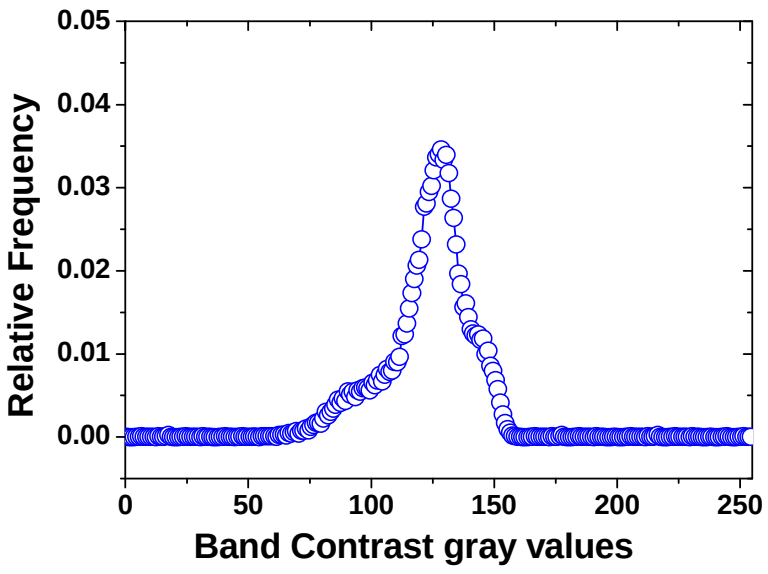

Fig. 10(a´´)

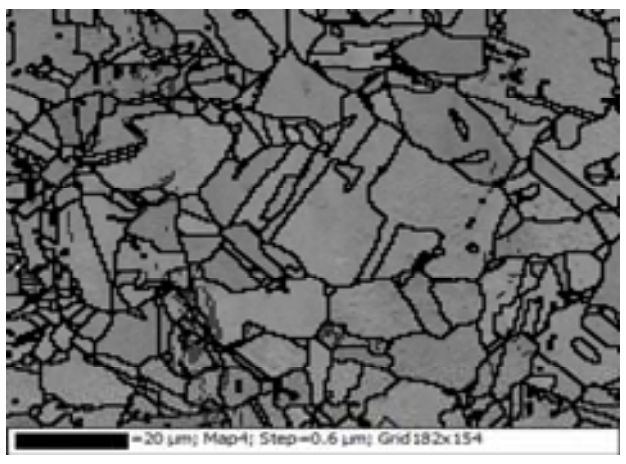

Fig. 10 (b)

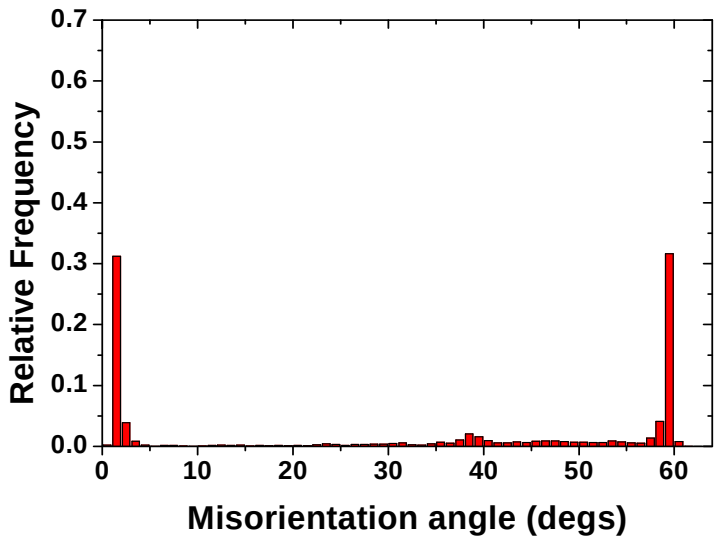

Fig. 10 (b´)

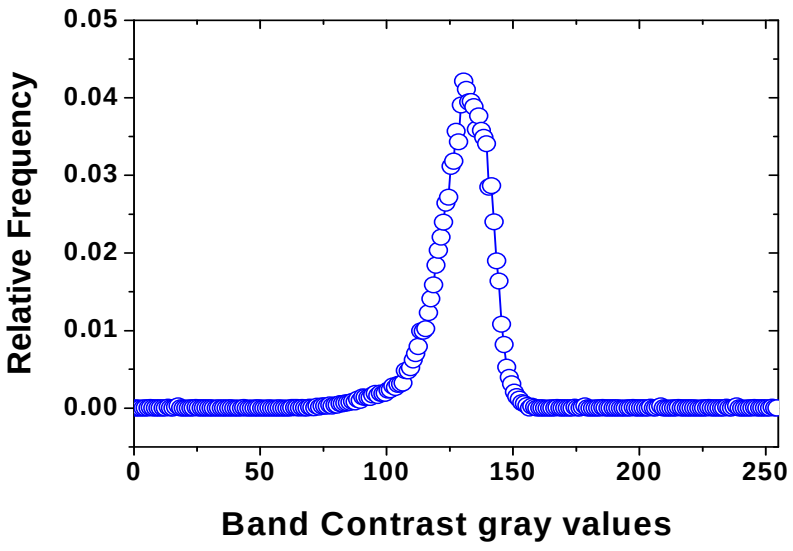

Fig. 10 (b´´)

Fig. 10. EBSD Micrograph of annealed samples (a) 50% rolled and (b) 80% rolled; Frequency distribution of misoriention angle of the annealed samples (a´) 50% rolled and (b´) 80% rolled ; Frequency distribution of band contrast of the annealed samples (a´´) 50% rolled and (b´´) 80% rolled

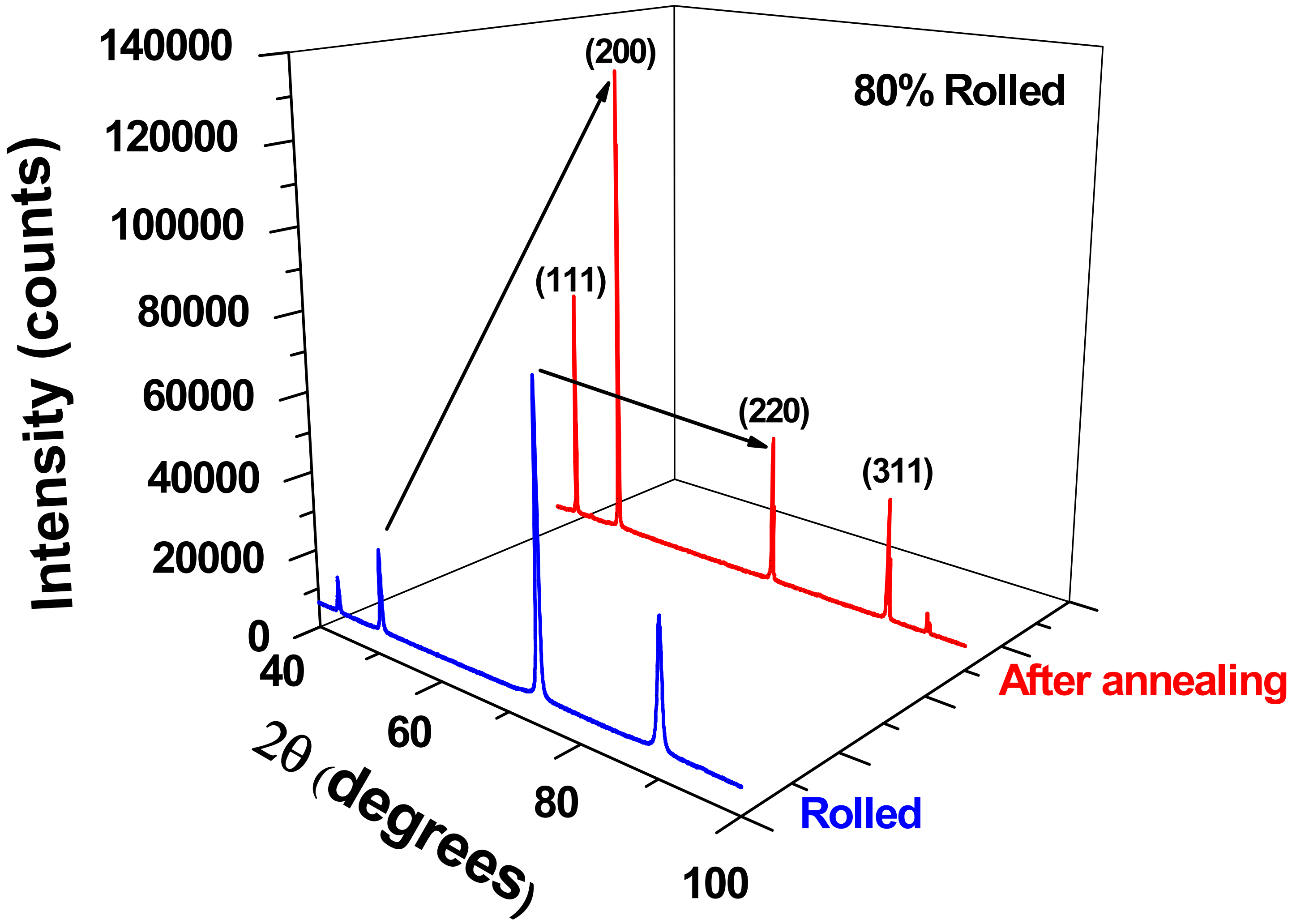

Fig. 11 X-ray diffraction patterns of the 80% rolled sample before and after annealing. (Arrows indicate the decrease in the (220) intensity and the increase in the (200) intensity).